\documentclass[twocolumn,reprint,amsmath,amssymb,floatfix,prl,preprintnumbers,footinbib, showpacs]{revtex4-1}
\newlength{\stdwidth}
\usepackage[dvips]{graphicx}

\usepackage{bbm}
\usepackage{amsfonts}
\usepackage{amsmath}
\usepackage{graphicx}
\usepackage[utf8]{inputenc}

\usepackage[usenames,dvipsnames]{xcolor}
\usepackage[colorlinks=true, linkcolor=RoyalBlue, citecolor=RoyalBlue]{hyperref}

\newcommand{\subref}[2]{\ref{#1}{\color{RoyalBlue}({#2})}}

\newcommand{\ci}{\mathrm{i}}
\newcommand{\e}{\mathrm{e}}
\newcommand{\eref}[1]{(\ref{#1})}
\newcommand{\textfrac}[2]{{#1}/{#2}}

\usepackage[normalem]{ulem}
\usepackage{color}

\begin{document}
\title{
Two-dimensional topological insulator edge state backscattering by dephasing 
}

\author{Sven Essert}
\affiliation{Institut f\"ur Theoretische Physik, Universit\"at Regensburg, D-93040 Regensburg, Germany}

\author{Viktor Krueckl}
\affiliation{Institut f\"ur Theoretische Physik, Universit\"at Regensburg, D-93040 Regensburg, Germany}

\author{Klaus Richter}
\affiliation{Institut f\"ur Theoretische Physik, Universit\"at Regensburg, D-93040 Regensburg, Germany}

\date{\today}

\begin{abstract}
To understand the seemingly absent temperature dependence in the conductance of two-dimensional topological insulator 
edge states, we perform a numerical study which identifies the quantitative influence of the combined effect of 
dephasing and elastic scattering in charge puddles close to the edges. We show that this mechanism may be responsible 
for the experimental signatures in HgTe/CdTe quantum wells if the puddles in the samples are large and weakly coupled
to the sample edges. We propose experiments on artificial puddles which allow to verify this hypothesis and to extract 
the real dephasing time scale using our predictions.
In addition, we present a new method to include the effect of dephasing in wave-packet-time-evolution algorithms.
\end{abstract}


\maketitle

The discovery of two-dimensional topological insulators, 2d-TIs, stirred up hope for potential applications
using their ability to host reflectionless one-dimensional transport. In practice,
however, the transport along the edges of the state-of-the-art realizations of 2d-TIs
exhibits a length-dependent resistance which increases above the expected
quantized value in samples significantly larger than $1~\mu \mathrm{m}$.
Moreover, most notably, the edge state resistance seems to exhibit no observable temperature dependence in a wide
temperature range ($T \approx 30\,\mathrm{mK}-30\,\mathrm{K}$) \cite{Spanton2014}. This seems to be hard to
reconcile with the notion that the backscattering on 2d-TI edges is due to inelastic processes. Still, this behavior 
has been been frequently observed in various recent experiments on 2d-TIs, both based on 
HgTe/CdTe~\cite{Gusev2014,Grabecki2013,Nowack2013} and InAs/GaSb samples~\cite{Spanton2014,Du2015}.
Understanding this apparently generic feature is important for the field, as it might show up ways to improve the edge
transport in existing material systems. 

As coherent elastic backscattering is symmetry forbidden at 2d-TI edges \cite{Kane2005a}, there have been theory
studies considering alternative backscattering mechanisms. Besides the possibility for elastic backscattering
by magnetic impurities which explicitly break time-reversal symmetry \cite{Maciejko2009,Tanaka2011,Cheianov2013}, they mainly focused on inelastic backscattering by 
Coulomb interaction or phonons on the edge \cite{Wu2006,Xu2006,Strom2010,Lezmy2012,Crepin2012,Budich2012,Schmidt2012,Kainaris2014,Geissler2014} as well as Coulomb backscattering in quantum dots 
which are expected to appear naturally along the edges of the currently used material systems (HgTe/CdTe and InAs/GaSb quantum wells) 
due to trapped charges at the gate insulator interface \cite{Vayrynen2013,Vayrynen2014}.

In addition, there have been a few proposals in which the inelastic interaction is mainly held responsible 
for breaking the electron phase coherence while the backscattering is then attributed to elastic
scattering. 
For a dephasing process which does not explicitly flip
the spin, it was found that one does not expect backscattering along a clean edge of a 2d-TI ribbon, as backscattering
requires a full spin flip in this setup~\cite{Jiang2009}. This changes if a quantum dot---in which
the extended states are naturally spin mixed due to the spin-orbit coupling---is present along or close to the edge.
In this setup, even a decoherence mechanism
which conserves the average spin may lead to backscattering, as has been qualitatively shown in Ref.~\onlinecite{Roth2009}
where dephasing was mimicked by adding lattice sites with an imaginary self energy.

In this publication, we also address dephasing-induced backscattering. However, given the fact that the interesting $T$-independent edge state resistance has been consistently observed
in various systems and material classes, we do not attempt to assign it to a specific source for decoherence
along with a corresponding specific $T$-dependence of the dephasing/decoherence time. Instead, 
assuming that edge channels are coupled to nearby puddles due to spatial charge inhomogeneities \cite{Roth2009,Grabecki2013,Konig2013,Vayrynen2013,Vayrynen2014} we 
present a more general study which is aiming at a quantitative description of backscattering through the interplay
of dephasing and spin mixing in such quantum dots. 
This includes to consider puddle dwell times and couplings to the puddles, which are calculated
for realistic model Hamiltonians for the case of the HgTe/CdTe material system. Thereby,
we extract a range of dephasing times that would be compatible with
the measured temperature-independent resistance, and which is a function of the puddle dwell time. This dependence
of the conductance on the puddle properties could be checked in future experiments with artificially created puddles.

\begin{figure}[tbh]
\centering
\includegraphics[width=\stdwidth]{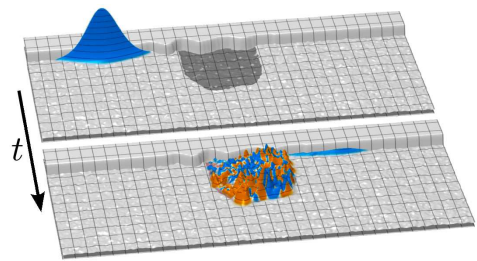}
\caption{(color online) Snapshots of the calculated wavefunction (the color encodes the spin) in the device geometry. A wave packet is approaching a puddle (top panel) in which it is fully spin-mixed by spin-orbit scattering (bottom). A video of the dynamics including dephasing is available online \cite{video}.
}
\label{figGeometry}
\end{figure}

\begin{figure*}[tbh]
\centering
\includegraphics[width=\textwidth]{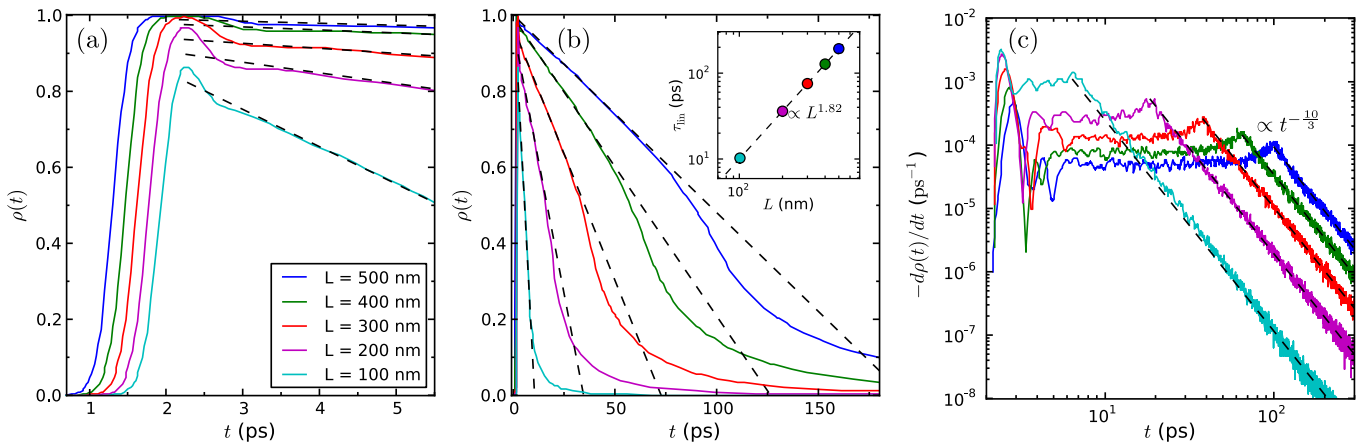}
\caption{(color online) Time-dependent probability density in a puddle of length $L$ close to the edge which is traversed by a wave packet starting on the edge close 
to the puddle at $t=0\,\mathrm{ps}$, see Fig.~\ref{figGeometry}. (a) Behavior for short times showing the steep rise in density from 
the wave packet entering the puddle and a small fraction which is leaving the puddle after traversing it ballistically. Most of
the density decays linearly (black dashed fit lines), for longer times shown in~(b). The inset shows the extracted linear decay coefficients $\tau_{\mathrm{lin}}$ (see main text)
from the fit
as a function of $L$ together with a power-law fit. (c) Log-log plot of the derivative of the density (the current)
demonstrating also the power-law decay of the density for long times.
}
\label{figPuddleOnEdge}
\end{figure*}

For our calculations, we use the Bernevig-Hughes-Zhang (BHZ) Hamiltonian~\cite{Bernevig2006} discretized on a square grid.
Intrinsic spin-orbit coupling of the Dresselhaus type is included, which mixes the two spin blocks~\cite{*[{}][{. We checked that Rashba-type SOI leads to similar effects.}] Konig2008}. 
We study the time evolution of a wave packet which is initially localized on the edge of a 2d-TI and
approaches a nearby puddle defined by a local electrostatic potential close to the edge, cf. Fig~\ref{figGeometry},
employing a method which has been shown to be well suited to describe the edge-state dynamics in HgTe/CdTe 2d-TIs \cite{Kramer2010,Krueckl2011a}. The
detailed calculation setup is described in detail in the Appendix.

From the time evolution of the wave packet, we calculate the time-dependent probability density $\rho(t)$ to find the charge carrier in the puddle, which includes
edge-puddle-coupling and lifetime effects. Impurity-configuration-averaged results of such calculations for various puddle sizes are shown in Fig.~\ref{figPuddleOnEdge}.
As can be seen from the short-time dynamics, Fig.~\subref{figPuddleOnEdge}{a}, the wave packet enters the cavity on a time scale
of a picosecond. A small fraction of the density, $2\operatorname{--}12\,\%$ depending on the puddle size, exits the puddle 
after a ballistic traversal. However, most of the density stays in the puddle
and decays only slowly with an almost constant outflow, visible as a linear density decay for intermediate times, cf. Fig.~\subref{figPuddleOnEdge}{b}. 
The shown density is scaled to the total density of the wave packet and the fact that values close to one are reached
reflects the forbidden backreflection on the clean edge: Most of the density has to couple into the puddle in this geometry.

As soon as the absolute density in the puddle falls below $\approx 0.5$, the time dependence of the density changes into a power law. 
This is best seen in Fig.~\subref{figPuddleOnEdge}{c}, which shows the negative time derivative of the density in a log-log plot. Here, one clearly
recognizes the linear density decay as a plateau with superimposed oscillations, which
can possibly be attributed to ballistic orbits in the puddle. This turns over into a $t^{-\frac{10}{3}}$-power-law decay for long times as indicated by the dashed
lines \footnote{This seems to be the right power for puddles larger than
$200\,\mathrm{nm}$ and implies $\rho(t)\propto t^{-\frac{7}{3}}$ for the density decay. For the $100\,\mathrm{nm}$ sized puddle, the best-fitting 
exponent would be slightly smaller, rather like $t^{-3.5}$, i.e., $\rho(t)\propto t^{-2.5}$.}.
Such power-law behavior is known \cite{Dittes1992} and similar results 
have already been obtained in wave-packet simulations on other material systems \cite{Zozoulenko2003}.

With only a few additional assumptions, the know\-ledge of the dwell time of the electrons in the quantum dot will allow us to estimate the effects 
of dephasing on its transport properties. The first one is that the decoherence disturbs the electron phase evolution but does not explicitly lead to spin flips, 
which is quite realistic for setups without magnetic fields and a very low density of magnetic 
impurities. Due to the strict spin-momentum locking of the TI edge states, this spin conservation implies that the dephasing process 
will not lead to backscattering as long as the electrons are propagating along the edges. In a quantum dot however, the spin
is fully mixed already after a very short time, due to the combined effect of impurity scattering and intrinsic spin-orbit coupling (see Fig.~\ref{figGeometry}). Thus, the
efficiency of backscattering due to dephasing will depend on the relation of the dephasing timescale and the electron lifetime
in the puddle. To make this quantitative, we assume that decoherence is statistically occuring with
independent events. We first additionally assume that each event leads to full phase loss (an assumption that we will
later refine). Thus, the electron density being in the puddle at the event
will leave it in a random direction in the subsequent dynamics.
Then, the average reflection can be estimated by
\begin{eqnarray}
R & = & \frac{1}{2\tau_{\phi}}\left[\int_{-\infty}^{t_{\mathrm{max}}}\! \! \! dt_{1}\rho(t_{1})e^{-\frac{t_{\mathrm{max}}-t_{1}}{\tau_{\phi}}}
                                   \!\!\!+\!\!\int_{t_{\mathrm{max}}}^{\infty}\! \! \! dt_{2}\rho(t_{2})e^{-\frac{t_{2}-t_{\mathrm{max}}}{\tau_{\phi}}}\right]\nonumber \\
 &  & -\frac{1}{2\tau_{\phi}^{2}\rho_\mathrm{max}}\int_{-\infty}^{t_{\mathrm{max}}}\! \! \! \! dt_{1}\int_{t_{\mathrm{max}}}^{\infty}\! \! \! \! dt_{2}\rho(t_{1})\rho(t_{2})e^{-\frac{t_{2}-t_{1}}{\tau_{\phi}}},\label{eq:bsformula}
\end{eqnarray}
where $\tau_{\phi}$ is the mean time between dephasing events and $t_\mathrm{max}$ is the
time at which the density $\rho(t)$ in the puddle attains its maximum $\rho_\mathrm{max}$. Equation~\eref{eq:bsformula} makes use of the piecewise monotonous structure of $\rho(t)$ and can be understood
as follows: Suppose that the first dephasing event occurs at a time $t_2 > t_\textrm{max}$ with the density already decaying monotonously. 
As each event is assumed to lead to full phase randomization, all subsequent
 events will not matter as the propagation in the puddle
is already fully random. Hence, the knowledge of the first event after reaching $\rho_{\mathrm{max}}$ is enough to determine 
the total backscattering probability for the time window of the decay. It can be calculated as an expectation value of
$\rho(t)$ with an exponential distribution for the dephasing event, describing the mean waiting time in a Poisson process. 
This main contribution enters Eq.~\eref{eq:bsformula} as the second term. The first term in Eq.~\eref{eq:bsformula}
describes the backscattering during the (monotonous) rise of $\rho(t)$. Here, the argument can be reversed
as the latest event will determine the amount of backscattered density in the time window of the rise.
The last term in Eq.~\eref{eq:bsformula} takes care of the double counting that occurs for events both on the rising and the falling
edge of $\rho(t)$.

For the shape of the density curves obtained from the numerical calculations, one can simplify the above expression. Since the rising edge
is very short compared to the decay, it suffices to consider only the second term of Eq.~\eref{eq:bsformula}. In addition, we
showed above that the subsequent decay is approximately linear. The power law decay can be omitted as long times are
exponentially suppressed in the integral.
We then find for the average transmission, using $\rho(t)\approx\rho_{\mathrm{max}}\left[1-\textfrac{(t-t_{\mathrm{max}})}{\tau_{\mathrm{lin}}}\right]$,
\begin{eqnarray}
T & \approx & 1-\frac{1}{2\tau_{\phi}}\int_{t_{\mathrm{max}}}^{t_{\mathrm{max}}+\tau_{\mathrm{lin}}}\! \! \! \! \! \! \! \! \! \! \! \! \! \! \! \! dt_{2}\rho_{\mathrm{max}}\left(1-\frac{t_{2}-t_{\mathrm{max}}}{\tau_{\mathrm{lin}}}\right)e^{-\frac{t_{2}-t_{\mathrm{max}}}{\tau_{\phi}}}\nonumber \\
 & = & 1-\frac{\rho_{\mathrm{max}}}{2}+\frac{\rho_{\mathrm{max}}\tau_{\phi}}{2\tau_{\mathrm{lin}}}\left(1-e^{-\frac{\tau_{\mathrm{lin}}}{\tau_{\phi}}}\right). \label{eq:simpl_linear}
\end{eqnarray}
Here $\tau_\mathrm{lin}$ is the time after which the puddle would be empty assuming a steady linear decay. The values for $\tau_\mathrm{lin}$ which were extracted from the linear
fits to $\rho(t)$ are plotted in a log-log plot in the inset of Fig.~\subref{figPuddleOnEdge}{b}. One finds that $\tau_\mathrm{lin}$ approximately scales like $L^{1.82}$.

\begin{figure}[tbh]
\centering
\includegraphics[width=\stdwidth]{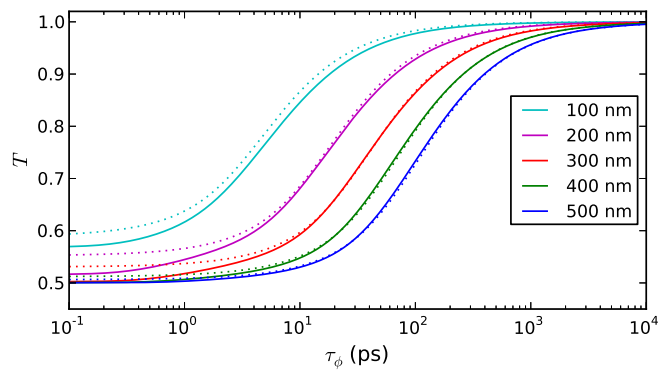}
\caption{(color online) Average transmission of a single quantum dot at the edge as a function of the dephasing time for different dot sizes. The full lines are evaluated from the full time-dependent
density, using Eq.~\eref{eq:bsformula}, while the dotted lines show the result from Eq.~\eref{eq:simpl_linear}, assuming a linear decay.
}
\label{figDecohfromModel}
\end{figure}
Using these formulae, we can estimate the average transmission probability for a single puddle of size $L$ for a given decoherence time $\tau_\phi$
as shown in Fig.~\ref{figDecohfromModel}. As anticipated, we find good agreement between the results from the full model, Eq.~\eref{eq:bsformula},
and the approximation with a linear fit, Eq.~\eref{eq:simpl_linear}. Both show a characteristic sigmoidal behavior with a saturation
at $T=1-\textfrac{\rho_{\mathrm{max}}}{2}$ for small $\tau_\phi$. For the smaller puddles, the extracted value for $\rho_{\mathrm{max}}$ is underestimated
for the linear fit, see Fig.~\subref{figPuddleOnEdge}{a}.

More generally, the value of $\rho_{\mathrm{max}}$ is not only given by the puddle size, but also by the coupling of the puddle
to the edge and it would therefore be strongly dependent on the distance of the puddle from the edge. This coupling will also influence the dephasing-time
scale below which one observes the nearly constant backscattering probability: With worse coupling, there will be less backscattering per puddle
but the saturation will already be reached earlier. The extracted cutoff times therefore refer to the ``perfect coupling limit'' meaning
that they may be interpreted as a lower bound for the cutoff time.

So far, the dephasing only entered the model a posteriori but was not included in the calculation of the electron dynamics. We found a way
to include it in the wave-packet dynamics calculation using an algorithm which is inspired by the concept of einselection \cite{Zurek2003}. It
treats dephasing in a rather general fashion and allows not only to vary the strength of single dephasing events but also
to study the influence of dephasing on the wave-packet dynamics itself. For details of the implementation, we refer to the Appendix.
The underlying interaction responsible for the dephasing is left unspecified and the
dephasing time remains as a parameter. Also, the implementation may not be applicable for all kinds of
environments but we think that it captures the important effects and expect similar results for other
implementations.

\begin{figure}[tbh]
\centering
\includegraphics[width=\stdwidth]{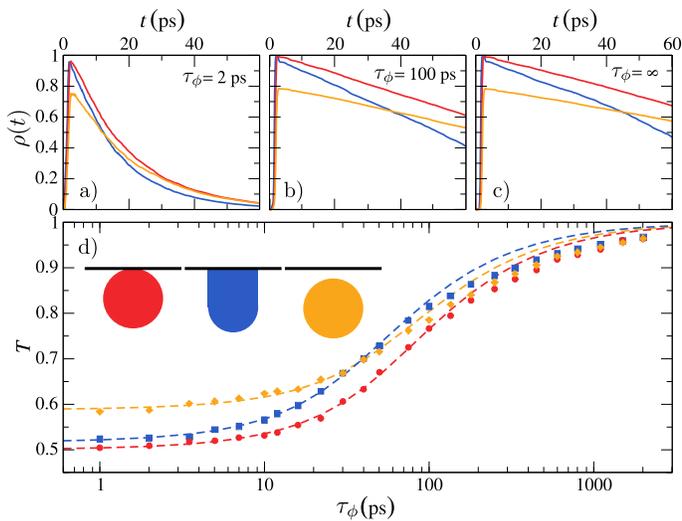}
\caption{(color online) (a)--(c) Time-dependent density in the puddle obtained including different effective dephasing times in the time evolution. The colors
correspond to different puddle geometries and couplings as shown in the inset in (d). (d) Averaged
transmission through the puddle as a function of the chosen dephasing time for the different puddle geometries. The dashed lines are fits to the numerical data (symbols) using Eq.~\eref{eq:simpl_linear}.
}
\label{figDecohNumerics}
\end{figure}

Our results for the total transmission of dynamical dephasing calculations are shown as symbols in Fig.~\subref{figDecohNumerics}{d}. Here, the dephasing was chosen to be less effective
such that one needs on average 2.75 events to achieve full dephasing (the time scale of the dephasing time axis corresponds to full dephasing). The dotted lines show a fit to the data
using Eq.~\eref{eq:simpl_linear} which expectedly agrees well for small dephasing times. In this limit, the time between the events is small compared to the dynamics of the system, thus
many small events lead to the same net result as one strong event. For longer dephasing times, however, in our refined calculation using partial dephasing, it comes into play that the dynamics of the system will be strongly influenced by the dephasing
as this effectively opens another exit channel for the puddle thus decreasing the electron lifetime. This can be nicely seen in the plots of the time-dependent density
for different dephasing times in Fig.~\subref{figDecohNumerics}{a--c}. In the long dephasing time limit, many frequent weak events lead to more backscattering than one rare strong event, as the second exit
channel (backscattering) will be (partially) opened already at an earlier time.

Fig.~\ref{figDecohNumerics} also shows the influence of different puddle geometries on the transmission and the system dynamics. The results match the expectation that puddles further
away from the edge will lead to less backscattering in the strong dephasing limit but will also saturate below a higher dephasing time cutoff.

With this, we can now relate to the experimental results on 2d-TI samples. So far, the measured resistance does not show
any observable temperature dependence in the experimentally accessible $T$-range 
($30\,\mathrm{mK}$--$30\,\mathrm{K}$)~\cite{Spanton2014}. This would be in line with our findings
if the dephasing times in the experimental samples were below the cutoff time even at the lowest temperature.
Then, an increase of the temperature (and a connected reduction of the dephasing time) would not lead to additional backscattering. 
To make a specific example, if the backscattering in the sample was mainly 
caused by $500\,\mathrm{nm}$ puddles which are well coupled to
the edges, this would imply that the dephasing time should be shorter than $\approx 10\,\mathrm{ps}$. 
For a rough comparison, from low-temperature magnetoconductance measurements on 2d HgTe quantum well samples, dephasing times of $\approx 70\,\mathrm{ps}$ were extracted \cite{Kozlov2013}. 

Thus,
given that the coupling to the puddles in the real samples is likely to be smaller, it might indeed be that the lack of observed temperature
dependence is mainly due to the inherently short dephasing times in these materials. We only consider puddles of a depth of $40\,\mathrm{meV}$
as we are limited to the range of validity of the BHZ-Hamiltonian. Deeper puddles, similar to larger puddles, would also shift the curves toward 
longer dephasing times and decrease the temperature above which the conductance is expected to be temperature independent.
The puddle depth and the connected change of electron density may also influence $\tau_\phi$, however, at least in the 2d-limit, recent experiments 
show controversial results whether an increasing density leads to an increased \cite{Kozlov2013} or a decreased \cite{Minkov2012} dephasing time.
Note that, in line with experiments, our model predicts reproducible conductance
fluctuations which are similar to UCFs as the coupling to the puddles is expected to be depending on gate voltage and magnetic field. However,
the conductance is not $T$-dependent as long as the dephasing time is below the cutoff.

To make a definite statement, one would have to experimentally check this hypothesis which could be done with experiments on artificially created
puddles. As there are experimental samples which show the quantization, it seems to be possible to create puddle-free samples in which one
could artificially introduce  puddles of varying size using a local gate. This should allow for 
 experimentally reproducing the calculated sigmoidal curve and extract the temperature-dependent dephasing time.

To summarize, we quantitatively studied the backscattering of 2d-TI edge states due to the interplay of dephasing and dynamical 
scattering in charge puddles. Our results suggest that the seemingly absent temperature dependence could be 
due to the saturation of backscattering at dephasing times which are short compared to the puddle lifetimes. 
One should be able to verify this hypothesis quantitatively with experiments on small ($\approx 100\,\mathrm{nm}$) artificial puddles, 
which, according to our study, should show a detectable temperature dependence at the experimentally available temperatures and 
from which one could extract the actual dephasing time.

In passing, we developed a scheme to
implement dephasing into wave-packet-time-evolution algorithms which is general and could be used in a wide range of scenarios. It is described
in detail in the Appendix. For the charge puddles in 2d-TIs, it would also allow for the explicit
inclusion of other sources of backscattering, like external magnetic fields, which we recently showed to have a strong influence on the backscattering due to puddles \cite{Essert2015a}.

This work was supported by DFG SPP 1666 and the ENB graduate school ``Topological Insulators''. We thank L.~Molenkamp and
 M.~Wimmer for fruitful discussions.


%

\vspace*{1cm}

\begin{appendix}
\section{Appendix}
\subsection{Model Setup}
In the main text, we show numerical calculations of an edge-state wave packet
which is scattered by an electrostatic puddle and extract the 
time-dependent probability density $\rho(t)$ of the wave packet being 
localized in the puddle.
We model the electronic structure of a HgTe/CdTe quantum well using 
the Bernevig-Hughes-Zhang Hamiltonian~\cite{Bernevig2006},
%
%
\begin{equation}
H=\left(\begin{array}{cc}
h(\mathbf{k}) & \begin{array}{cc}
0 & -\varDelta\\
\varDelta & 0\end{array}\\
\begin{array}{cc}
0 & \varDelta\\
-\varDelta & 0\end{array} & h^{*}(-\mathbf{k})\end{array}\right),
\label{bhzhamilton}
\end{equation}
%
%
with spin-subblock Hamiltonians
%
%
\begin{widetext}
\begin{equation}
h(\mathbf{k})=\left(\begin{array}{cc}
V(x)+M(x)-(B+D)\mathbf{k}^{2} & Ak_{+}\\
Ak_{-} & V(x)-M(x)+(B-D)\mathbf{k}^{2}\end{array}\right),
\label{eq:subblockhamil}
\end{equation}
\end{widetext}
%
%
where $k_{\pm}=k_{x}\pm i k_{y}$ and $\mathbf{k}^{2}=k_{x}^{2}+k_{y}^{2}$.
Intrinsic spin-orbit coupling of the Dresselhaus-type is included by the parameter~$\varDelta$ in Eq.~\eref{bhzhamilton},
which mixes the two spin blocks~\cite{Konig2008}.
We use an electrostatic potential $V(x)$ to model circular and stadium
shaped puddles and confine the states by the potential $M(x)$ to get
a quantum spin Hall edge state at the system boundary.
For the calculations presented in the manuscript, the potential strength
of the puddle was set to $40~{\rm meV}$ leading to hole-like states.

As initial state, we create a Gaussian edge-state wave packet $\psi(t_0)$,
which is assembled in reciprocal space along the boundary.
Throughout the manuscript, we use a Gaussian with a width in position space of
$\sigma = 90~{\rm nm}$. Also, we include only states  propagating towards the
electrostatic puddle, leading to a strongly spin-polarized wave packet.
We calculate the propagation of the wavefunction $\psi(t)$
 using a propagator based on an expansion
of the time-evolution operator in  Chebyshev Polynomials~\cite{Tal-Ezer1984}.
During the propagation of the wave packet $\psi(t)$, we integrate $\left|\psi(t)\right|^2$ over the
puddle region resulting in the time-dependent probability density $\rho(t)$.
In order to avoid effects of mesoscopic fluctuations, we 
average over 20 different configurations, which differ in
 a random impurity potential with an amplitude of
$5~{\rm meV}$ and a wall distortion of $20~{\rm nm}$.

\subsection{Propagation with dephasing}
In the following section, we will show how dephasing is included in the
numerical calculations presented in Fig.~~\ref{figDecohNumerics}.
The dephasing algorithm described here is inspired by the concept of einselection
(``environment-induced superselection'') pioneered by W. Zurek \cite{Zurek2003}, 
which is a mechanism proposed to understand the influence of dephasing on quantum systems
and, more generally, to explain the quantum-to-classical transition.
According to einselection, the interaction of an open system with its environment
leads to decoherence \footnote{we use decoherence and dephasing synonymously in this article},
which causes a decay of the quantum states of the system into an incoherent
mixture of so-called pointer states.
This strongly suppresses quantum interference effects between different pointer
states on the time scale of the dephasing time $\tau_\phi$.
The character of the set of pointer states, the pointer basis, heavily depends on
the environment and the coupling to it.
For example, for very weak coupling to the environment, the pointer basis coincides
with the set of energy eigenstates of the system.
However, in the case of an intermediate system-environment coupling based on a local interaction,
e.~g., in the case of electron-phonon or electron-electron coupling, it is
expected to be a set of states that is localized in phase space, i.~e.,
in position and momentum.

\begin{figure}[t]
\centering
\includegraphics[width=.9\stdwidth]{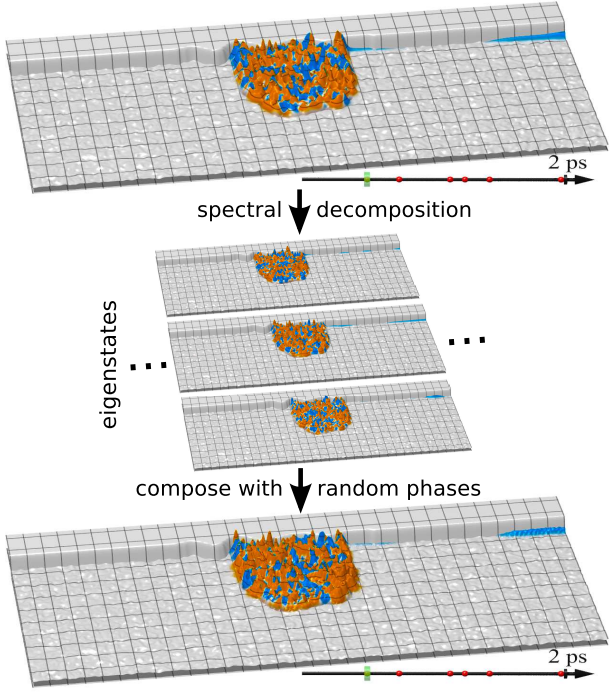}
\caption{The figure illustrates the workings of the dephasing algorithm. At a dephasing event, the current wavefunction---here showing a wave packet spread in 
an electrostatic puddle (the color encodes the spin)---is spectrally decomposed in a set of pseudo eigenstates and recomposed with random phases, leading to
a new dephased wavefunction.
A video using this type of dephasing is available online~\cite{video}.
}
\label{figDecoposition}
\end{figure}

As in the puddle lifetime calculations, we again employ a numerical time evolution based
on a single pure state, i.~e., we do not use a representation in terms of
density matrices, which would drastically increase the computational effort.
Still, we incorporate the dephasing-induced interference suppression by occasionally (partially)
randomizing the phases of the components of the state vector in a representation that
tries to faithfully mimic a decomposition in terms of the pointer basis.
This randomization is done at fixed event times $t_n$, which are sampled from an exponential
distribution with time constant $\tau_\mathrm{e}$, i.~e., we assume the events to be fully
uncorrelated (Poisson process).
In between these events, the propagation is done fully coherently using the
propagator based on a polynomial expansion mentioned above.
The decomposition and the subsequent randomization is done in the following way: At the
time of an event $t_n$, we extract a set of pseudo eigenstates,
%
%
\begin{equation}
\phi_m \propto \int_{t_n-\Delta t}^{t_n+\Delta t}  \psi(t)\,\e^{\ci E_m t/\hbar} {\rm d}t,
\end{equation}
%
%
at the energies $E_m = \{0\,{\rm meV}, \pm 1.5\,{\rm meV}, ... \pm 7.5\,{\rm meV} \}$
using a short-time propagation of the wave packet $\psi(t)$ around the time $t_n$.
These states fulfill the requirement that they are both local in energy as well
as in position space (as they are extracted from a propagation over a finite time interval).
The degree of localization in position space can by tuned by the propagation time $\Delta t$ \footnote{For 
$\Delta t\rightarrow \infty$ and a matching set of energies, this approach yields the expected pointer 
basis in the limit of very weak coupling to the environment: the exact energy eigenstates of the system.},
which we choose $\Delta t\sim 1\,\mathrm{ps}$.
We use these states $\phi_m$ to spectrally decompose the current state before the event $\psi(t_n)$, 
as sketched in Fig.~\ref{figDecoposition}.
%
%
%
%
For this, we first orthogonalize them with a Gram-Schmidt process,
leading to the set $\tilde \phi_m$.
Then, we calculate the weights $a_m$, which can be used to express the current state as the decomposition
%
%
\begin{equation}
\psi(t_n) = \sum_{m} a_m  \tilde \phi_m + \Delta \psi \text{.}
\end{equation}
%
%
Since the set of 11 states is not sufficient to describe the wave packet
$\psi(t_n)$ exactly, we also allow for account a small residual part $\Delta \psi$.
In this decomposed representation, the dephasing can be easily added by modifying the set 
of amplitudes to
%
%
\begin{equation}
\tilde a_m = a_m  \exp(\ci \pi ~ {\rm rand} ),
\end{equation}
%
%
with the random numbers ${\rm rand} \in [-Q, Q]$.
Subsequently, we create the new wave packet
%
%
\begin{equation}
\psi(t_n+\epsilon) = \sum_{m} \tilde a_m  \tilde \phi_m + \Delta \psi \text{,}
\end{equation}
%
%
which is used in the following propagation.
With this dephasing algorithm, we are able to calculate the time evolution in 
arbitrary mesoscopic systems.
It roughly conserves the energy, the spatial extent, as well as the total spin of the wave packet, which
is what one would expect from the interaction with a spin-unpolarized (non-magnetic) environment.
The strength of a single dephasing event can be tuned by the degree of randomization of the phase
in each event, i.~e., by the parameter $Q$. 
This also makes the regime of weak but frequent dephasing accessible (compared to the regime
of rare and very strong dephasing which allows a simpler model treatment discussed
in the first part of the manuscript).
However, to compare with the experimentally accessible dephasing time $\tau_\phi$, which is
the time for full phase coherence loss, one has to
find the relation between the time constant of the events $\tau_\mathrm{e}$ and the
time for full dephasing $\tau_\phi$.
This can be done in the following way: The average correlation of a state in the pointer basis $|\psi^\mathrm{P}\rangle$ with itself after $n$ events is given by,
\begin{align}
&\iiint_{-Q}^{Q}dq_{1}\dots dq_{n}e^{i\pi\sum_{j=1}^{n}q_{j}}\nonumber \\
=\, &\frac{1}{\left(2Q\right)^{n}}\iiint_{-Q}^{Q}dq_{1}\dotsm dq_{n}\cos\left(\pi\sum_{j=1}^{n}q_{j}\right)\nonumber \\
=\, &\left(\frac{1}{2Q}\int_{-Q}^{Q}dq\cos\left(\pi q\right)\right)^{n}=\left(\frac{\sin\left(\pi Q\right)}{\pi Q}\right)^{n}.
\end{align}
With the chance for $n$ events occurring after time $t$ in a Poisson process,
\begin{equation}
P(n)=\frac{t^{n}}{\tau_{\mathrm{e}}^{n}}\frac{1}{n!}e^{-\frac{t}{\tau_{\mathrm{e}}}},
\end{equation}
one can evaluate the average decay of autocorrelation after time $t$,
\begin{align}
&\frac{\langle\psi^{\mathrm{P}}(0)|\psi^{\mathrm{P}}(t)\rangle}{\langle\psi^{\mathrm{P}}(0)|\psi^{\mathrm{P}}(t)\rangle_{\mathrm{coherent}}}\nonumber \\
=\, &\sum_{n=0}^{\infty}P(n)\left(\frac{\sin\left(\pi Q\right)}{\pi Q}\right)^{n}\nonumber \\
=\, &\sum_{n=0}^{\infty}\frac{1}{n!}\left(\frac{t}{\tau_{\mathrm{e}}}\frac{\sin\left(\pi Q\right)}{\pi Q}\right)^{n}e^{-\frac{t}{\tau_{\mathrm{e}}}}\nonumber \\
=\, &\exp\left[\frac{t}{\tau_{\mathrm{e}}}\frac{\sin\left(\pi Q\right)}{\pi Q}-\frac{t}{\tau_{\mathrm{e}}}\right]\nonumber \\
=\, &\exp\left[-\frac{t}{\tau_{\mathrm{e}}}\left(1-\frac{\sin\left(\pi Q\right)}{\pi Q}\right)\right]=\exp\left[-\frac{t}{\tau_{\phi}}\right],
\end{align}
with
\begin{equation}
\tau_{\phi}=\frac{\pi Q}{\pi Q-\sin\left(\pi Q\right)}\tau_{\mathrm{e}},
\end{equation}
yielding the desired relation between decoherence time $\tau_\phi$ and the time constant of the events $\tau_\mathrm{e}$. 
To model the frequent but weak dephasing regime, we use $Q=\frac{1}{2}$, thus, $\tau_\phi=\pi/(\pi-2)\tau_\mathrm{e}\approx 2.75\,\tau_\mathrm{e}$.

In Fig.~\ref{figDecohNumerics}, we apply the dephasing scheme on three different electrostatic
puddle shapes to show how the coupling between the puddle and the edge states affects the total 
spin randomization of the puddle in presence of dephasing.
A video of a sample propagation can be also found online \cite{video}.

\end{appendix}

\end{document}